\documentclass[preprint,aps,nofootinbib,newabstract]{revtex4}

\usepackage{graphicx}
\usepackage{latexsym}

\newcommand{\lsim}{\mathrel{\mathop{\kern 0pt \rlap
  {\raise.2ex\hbox{$<$}}}
  \lower.9ex\hbox{\kern-.190em $\sim$}}}
\newcommand{\gsim}{\mathrel{\mathop{\kern 0pt \rlap
  {\raise.2ex\hbox{$>$}}}
  \lower.9ex\hbox{\kern-.190em $\sim$}}}
\newcommand{\beq}     {\begin{equation}}
\newcommand{\eeq}     {\end{equation}}
\newcommand{\bea}     {\begin{eqnarray}}
\newcommand{\eea}     {\end{eqnarray}}

\newcommand{\gm}      {\gamma}
\newcommand{\Gam}      {\Gamma}

\newcommand{\M}       {{\mathcal M}}

\newcommand{\no}      {\nonumber}
\newcommand{\hats}      {\hat{s}}

\newcommand{\nc}{\newcommand}
\nc{\postscript}[2]
{\setlength{\epsfxsize}{#2\hsize}\centerline{\epsfbox{#1}}}
\nc{\non}{\nonumber}
\nc{\hc}{\hbox {h.c.}} \nc{\re}{\hbox {Re}} 
\nc{\mev}{\hbox {MeV}} \nc{\gev}{\;\hbox {GeV}} \nc{\tev}{\;\hbox {TeV}}
\def\lsim{\mathrel{\raise.3ex\hbox{$<$\kern-.75em\lower1ex\hbox{$\sim$}}}}
\def\gsim{\mathrel{\raise.3ex\hbox{$>$\kern-.75em\lower1ex\hbox{$\sim$}}}}

\nc{\Lsp}{\;\;\;\;\;\;\;\;\;\;}  \nc{\LLLsp}{\lspace \lspace}
\nc{\lsp}{\;\;\;\;\;\;}
\nc{\spac}{\;\;\;}
\nc{\noi}{\noindent}
\nc{\baa}{\begin{array}}      \nc{\eaa}{\end{array}}
\nc{\bit}{\begin{itemize}}    \nc{\eit}{\end{itemize}}
\nc{\ben}{\begin{enumerate}}  \nc{\een}{\end{enumerate}}
\nc{\bce}{\begin{center}}     \nc{\ece}{\end{center}}

\def\Hhat{\widehat H}

\def\gam{\gamma}

\def\lwh{\widehat\Lambda_W}

\def\lphi{\Lambda_\phi}

\def\hbar{\overline h}
\def\grav{h_{\mu\nu}^{(n)}}
\def\lm{\lambda}
\def\Lm{\Lambda}
\def\mpl{M_{\rm Pl}}
\def\ifmath#1{\relax\ifmmode #1\else $#1$\fi}

\begin{document}

\renewcommand{\thepage}{-- \arabic{page} --}
\def\mib#1{\mbox{\boldmath $#1$}}
\def\bra#1{\langle #1 |}      \def\ket#1{|#1\rangle}
\def\vev#1{\langle #1\rangle} \def\dps{\displaystyle}
\nc{\tb}{\stackrel{{\scriptscriptstyle (-)}}{t}}
\nc{\bb}{\stackrel{{\scriptscriptstyle (-)}}{b}}
\nc{\fb}{\stackrel{{\scriptscriptstyle (-)}}{f}}
\nc{\pp}{\gamma \gamma}
\nc{\pptt}{\pp \to \ttbar}
\nc{\barh}{\overline{h}}
   \def\thebibliography#1{\centerline{REFERENCES}
     \list{[\arabic{enumi}]}{\settowidth\labelwidth{[#1]}\leftmargin
     \labelwidth\advance\leftmargin\labelsep\usecounter{enumi}}
     \def\newblock{\hskip .11em plus .33em minus -.07em}\sloppy
     \clubpenalty4000\widowpenalty4000\sfcode`\.=1000\relax}\let
     \endthebibliography=\endlist
   \def\sec#1{\addtocounter{section}{1}\section*{\hspace*{-0.72cm}
     \normalsize\bf\arabic{section}.$\;$#1}\vspace*{-0.3cm}}
\preprint{ \hbox{\bf hep-ph/yymmnn}}

\vspace*{2cm}

\title{Probing the radion-Higgs mixing at photon colliders}

\author{
Kingman Cheung\footnote{cheung@phys.nthu.edu.tw}}
\affiliation{Department of Physics and NCTS, National
Tsing Hua University, Hsinchu, Taiwan}
\author{
C.~S. Kim\footnote{cskim@yonsei.ac.kr,~~ JSPS Fellow}
}
\affiliation{Department of Physics, Yonsei University, Seoul 120-749, Korea\\
Faculty of Sciences, Hiroshima University, Higashi-Hiroshima, Japan}
\author{
Jeonghyeon Song\footnote{jhsong@konkuk.ac.kr}}
\affiliation{Department of Physics, Konkuk University,
                   Seoul 143-701, Korea}

\begin{abstract}

\noindent In the Randall-Sundrum model, the radion-Higgs mixing is
weakly suppressed by the effective electroweak scale. A novel
feature of the gravity-scalar mixing would be a sizable three-point
vertex of $\grav$-$h$-$\phi$. We explore the potential of photon
colliders, achieved by the laser backscattering technique, in
probing the radion-Higgs mixing via the associated production of the
radion with the Higgs boson. The advantage of photon colliders is
the capability of adjusting the polarization of the incoming photons
such that the signal of the spin-2 graviton exchange can be largely
enhanced. The enhancement factor is shown to be about 5, except for
small-$\xi$ region. We also study the corresponding backgrounds
step-by-step in detail.
\end{abstract}

\maketitle

\section{Introduction}

The standard model (SM) has been
extraordinarily successful up to now
in explaining all experimental data on the electroweak
interactions
of the gauge bosons and fermions.  However,
the master piece of the SM,
the Higgs boson, which is responsible for the electroweak
symmetry breaking, still awaits
experimental discovery\,\cite{Higgs}.
The direct search has excluded the SM Higgs boson mass below about
114 GeV\,\cite{higgs}, while the indirect evidences from the
electroweak precision data put an upper bound for a SM Higgs of 208
GeV at 95\% C.L. \,\cite{lepew}. The precision measurements and the
direct searches are getting into the situation that they begin to
contradict each other. There have been various studies which can
ease the situation. 
One possibility is that the SM Higgs boson mixes with another scalar boson
such that the Higgs branching ratio into $b\bar b$ becomes smaller and
thus escapes the limit of direct search so far.
Disentangling the nature of this new scalar state
is very significant and challenging.

It has been pointed out that the radion of the
Randall and Sundrum (RS) model \cite{RS} can play the role of such a
scalar boson.
The RS model consists of
an additional spatial dimension of a $S^1/Z_2$ orbifold introduced
with two 3-branes at the fixed
points.  A geometrical suppression factor, called the warp factor,
explains the huge hierarchy between the electroweak and Planck scale
with moderate values of model parameters.
The presence of a radion, the quantum degree with respect to the fluctuation
of the brane separation,
naturally emerges from the stabilizing process\,\cite{GW,Csaki-cosmology}.
As various stabilization mechanisms suggest,
the radion is generically much lighter
than the Kaluza-Klein (KK) states of any bulk field.
In the literature,
phenomenological aspects of the radion
have been studied such as
its decay modes\,\cite{Ko,Wells},
its effects
on the electroweak precision observations\,\cite{Csaki-EW},
and its signatures
at present and future colliders\,\cite{collider}.

The radion-Higgs mixing is originated
from the gravity-scalar mixing term,
$\xi R(g_{\rm vis}) \widehat{H}^\dagger \widehat{H}$,
where $R(g_{\rm vis})$ is the Ricci scalar of the induced metric
$g_{\rm vis}^{\mu\nu}$, and $\widehat{H}$ is the Higgs field in the
five-dimensional context.
It has been shown that the radion-Higgs mixing can induce significant
deviations to the properties of the SM Higgs
boson\,\cite{Datta-HR-LHC,Han-unitarity,Hewett:2002nk,Gunion,toharia}.
A complementary way to probe the radion-Higgs mixing
is the direct search for the new couplings
exclusively allowed with a non-zero mixing parameter $\xi$.
One good example is the tri-linear vertex among
the KK graviton field $h_{\mu\nu}^{(n)}$, the Higgs boson $h$, and the radion
$\phi$.
In Refs.~\cite{ours,ours2},
we have shown that
probing the vertex $h_{\mu\nu}^{(n)}$-$h$-$\phi$
through the $h \phi$ production at $e^+ e^-$ colliders and hadronic colliders
can provide very useful information on the radion-Higgs mixing,
irrespective of the mass spectrum of the Higgs boson and the radion.

In this work, we turn to photon colliders achieved by the laser
backscattering technique \cite{Ginzburg}.
The process that we investigate is
\begin{equation}
\gamma \gamma \to h^{(n)}_{\mu\nu} \to h \phi~,
\end{equation}
where $h^{(n)}_{\mu\nu}$ denotes the $n$-th KK state of the RS graviton.
Since the polarization of
incoming photons can be adjusted by tuning the polarization of the
electron or positron beam and the laser beam,
the signal can be largely enhanced because the
exchanged graviton is a spin-2 particle.
This is the biggest advantage
of photon colliders in this regard.
The observation of the rare decay of a KK graviton into  $h \phi$
is then the direct and exclusive signal of the radion-Higgs mixing.
In addition, the characteristic
angular distribution could reveal the exchange of massive spin-2 KK
gravitons.

This paper is organized as follows.
In Section II, we calculate the production cross section of
$\gamma\gamma \to h^{(n)}_{\mu\nu} \to h \,\phi$ folded with the photon
luminosity function.
Section III
deals with the feasibility of detecting the $h \phi$ final states
by considering specific decay channels of the Higgs boson and the radion.
We summarize at the end of Section III.
Note that we shall use $G^{(n)}$ or $h^{(n)}_{\mu\nu}$ to
denote the $n$-th KK graviton state interchangeably.

\section{Calculation of $\gamma\gamma \to h^{(n)}_{\mu\nu} \to h \phi$}

The RS scenario is based on a five-dimensional spacetime of a
$S^1/Z_2$ orbifold which has the finite size of $b_0$.
The warped factor, $\Omega_0=e^{-m_0 b_0/2}$,
with a moderate value of $m_0 b_0/2 \simeq 35$
can solve the gauge hierarchy problem.
In terms of the KK graviton field $h^{(n)}_{\mu\nu}$ and the canonically
normalized
radion field $\phi_0$,
the four-dimensional effective
Lagrangian is then
\beq
{\cal L}=
-\frac{\phi_0}{\Lambda_\phi}
T_\mu^\mu -\frac{1}{\lwh} T^{\mu\nu}(x) \sum_{n=1}^\infty
h^{(n)}_{\mu\nu}(x) \,,
\eeq
where $\Lambda_\phi$ is the vacuum expectation value (VEV) of the radion
field, $T^\mu_\mu$ is the trace of the symmetric energy-momentum
tensor $T^{\mu\nu}$, and $\lwh $ is the effective electroweak scale.
Both effective interactions are suppressed by the electroweak scale,
not by the Planck scale.
The gravity-scalar
mixing term of
$S_\xi=\xi \int d^4 x \sqrt{g_{\rm vis}}\,R(g_{\rm vis})
\Hhat^\dagger \Hhat$~\cite{Wells,Gunion}
is allowed as it respects all the SM symmetries and Poincare invariance.
Here $g_{\rm vis}$ is the induced metric
on the visible brane, $R(g_{\rm vis})$ is the Ricci scalar,
$H_0=\Omega_0 \Hhat$, and the dimensionless parameter
$\xi$ of order one denotes the size of the mixing term.
This $\xi$-term mixes the $h_0$ and $\phi_0$ fields
into the mass eigenstates of $h$ and $\phi$ fields~\cite{Gunion}:
\beq
\label{matrix}
\left(
\begin{array}{c}
  h_0 \\
  \phi_0 \\
\end{array}
\right)
=
\left(
\begin{array}{cc}
  d & c \\
  b & a \\
\end{array}
\right)
\left(\begin{array}{c}
  h \\
  \phi \\
\end{array}
\right) \,.
\eeq
We refer the detailed expressions for $a$, $b$, $c$, and $d$
to Ref.~\cite{ours2}.

All phenomenological signatures of the RS model
are then determined by
five parameters of
\beq
\label{parameter}
\xi,\quad \lphi,\quad \frac{m_0}{\mpl},\quad m_\phi,\quad
m_h
\,,
\eeq
which in turn determine $\lwh={\lphi}/\sqrt{3}$ and
KK graviton masses $m_{G^{(n)}}=x_n {m_0}{\lwh}/({\mpl} {\sqrt{2}})$
with $x_n$ being the $n$-th root of the first order Bessel function of the
first kind.
The ratio $m_0/\mpl$
is assumed in $[0.01 ,~ 0.1]$
to avoid too large bulk curvature\,\cite{Hewett-bulk-gauge}.
In what follows, we fix the ratio $m_0/\mpl=0.1$.
The $\Lm_\phi$ or $\lwh$ is constrained
by the Tevatron Run I data of Drell-Yan process
and by the electroweak precision data:
For $m_0/\mpl=0.1$,
$m_G^{(1)}\gsim 500$ GeV yields $\lphi\gsim 3.2$ TeV\,\cite{RS-onoff}.
Therefore, we consider the case of $\lphi=3.5$ TeV and $m_0/\mpl = 0.1$,
of which
the effect of radion on the oblique parameters
is small\,\cite{Csaki-EW}.
Then, the first KK graviton mass is about 547.5 GeV.
The radion mass is expected to be light
as one of the simplest stabilization mechanisms
predicts $m_{\phi_0} \sim \lwh/40$\,\cite{GW}.
In addition, the Higgs boson mass is set to be $120$ GeV through out the paper.

%
The gravity-scalar mixing
modifies the couplings among the $h$, $\phi$ and $\grav$.
In particular, a non-zero $\xi$ gives rise to
new tri-linear vertices of
\beq
\label{4vertices}
\grav\,\mbox{-}\,h\,\mbox{-}\,\phi,\quad
\grav\,\mbox{-}\,\phi\,\mbox{-}\,\phi,\quad
h\,\mbox{-}\,\phi\,\mbox{-}\,\phi, \quad
\phi\,\mbox{-}\,\phi\,\mbox{-}\,\phi
\,.
\eeq
Due to
the suppressed coupling of a photon with a Higgs boson or a radion,
the $\gamma\gamma$ collider is expected to access the
$\grav$-$h$-$\phi$ or $\grav$-$\phi$-$\phi$ vertex directly.
In addition, the coupling strength of $\grav$-$\phi$-$\phi$
is much smaller than
that of $\grav$-$h$-$\phi$,
by a factor of $\gam \equiv v/\lphi \ll 1$.
Here $v$ is the VEV of the Higgs boson, which is 246 GeV.
Therefore, the channel $\grav \to h \phi$ is the most effective in
probing the radion-Higgs mixing,
with the vertex denoted by
\begin{equation}
\langle \, h \, | \,\grav \, | \,\phi \rangle \equiv i \hat{g}_{G
h\phi} \frac{2 k_{1\mu}k_{2\nu} }{\lwh}~,
\end{equation}
where $ \hat{g}_{G h\phi} = 6\gam\xi\left[a(\gam b+d)+bc\right]+cd$
and $k_{1,2}$ is the four-momentum of the scalar particles.
Then, the partial decay width of $\grav \to h \phi$ is given as
\beq
\Gamma(h^{(n)}_{\mu\nu} \to h \phi  ) =
\frac{\hat{g}^2_{G h \phi}}{ 240\pi}\,
         \frac{m_G^3}{\hat \Lambda_W^2} \, \beta
  \left[ 1 - \left( \sqrt{\mu_{h G}}+\sqrt{\mu_{\phi G}} \right )^2
 \right ]^2 \,
  \left[ 1 - \left( \sqrt{\mu_{h G}}-\sqrt{\mu_{\phi G}} \right )^2 \right ]^2,
\eeq
where $\mu_{xy} = (m_x/m_y)^2$,
$\beta=\lambda^{1/2}(1, \mu_{\phi G}, \mu_{h G})$,
and $\lambda(a,b,c)=a^2+b^2+c^2-2 a b -2 a c-2 b c$.
In Ref.~\cite{ours2},
it was shown that the branching ratio
${\rm Br} (h^{(n)}_{\mu\nu} \to h \phi, \phi \phi)$,
which would vanish in the limit $\xi \to 0$,
is of the order of $\mathcal{O} (10^{-3})$.

\begin{figure}[t!]
\centering
\includegraphics[width=4in]{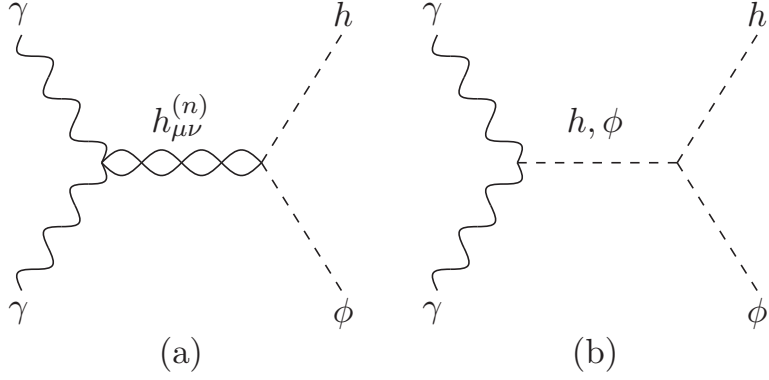}
\caption{\label{fig1} \small
Feynman diagrams for the process $\gamma\gamma
\to h \phi$.}
\end{figure}

For the process of
\beq
\label{rrhr}
\gamma(q_1,\lm_1) + \gamma(q_2,\lm_2)  \to h(k_1)+ \phi(k_2)
\,,
\eeq
the Feynman diagrams are shown in Fig. \ref{fig1}.
Here $\lm_{1,2}$ is the polarization of the high energy photons.
The helicity amplitudes
$\M_{\lm_1 \lm_2}$ including the $h$ and $\phi$ mediation are
\bea
\label{eq:helicityamp++}
\M_{++} &=& -
\frac{\hats}{2 v \lphi} \left( \hat{c}_{\gm \gm h} \hat{g}_h \mathcal{D}_h +
\hat{c}_{\gm\gm \phi} \hat{g}_\phi \mathcal{D}_\phi \right),
\\ \label{eq:helicityamp+-}
\M_{+-} &=& -\frac{\hats }{\lwh^2}
\,\lm(1,m_h^2/\hats,m_\phi^2/\hats)
\hat{g}_{G\phi
h}\mathcal{D}_G \sin^2\theta^* \,,
\eea
where $\hats = (q_1+q_2)^2$,
$\theta^*$ is the scattering angle of the Higgs boson in the $\gm\gm$ c.m.
frame, $\hat{c}_{\gm\gm h}$ and $\hat{c}_{\gm \gm\phi}$ are
\bea
\hat{c}_{\gm \gm h}
 &=&
 - \frac{\alpha }{2\pi }
 \left[
 (d + \gm b) \sum_i e_i^2 N_c^i F_i(4 m_i^2/m_h^2) +\frac{11}{3} \gm b
 \right]
 \,, \\ \no
\hat{c}_{\gm \gm \phi}
 &=&
 - \frac{\alpha }{2\pi }
 \left[
 (c + \gm a) \sum_i e_i^2 N_c^i F_i(4 m_i^2/m_\phi^2) +\frac{11}{3} \gm a
 \right].
\eea
We refer the expressions for $F_{1/2}$, $F_1$, and $\hat{g}_{h,\phi}$
to Ref.~\cite{ours2}.
The propagator factors of the KK-graviton, the Higgs boson, and
the radion are given by
\beq
\label{eq:DG}
{\mathcal D}_G =\sum_{n=1}^\infty
\frac{\hats}{\hats - m^2_{G^{(n)}}+ i m_{G^{(n)}}\Gam_{G^{(n)}} } \,,\quad
{\mathcal D}_{h,\phi} =\frac{\hats}{\hats-m_{h,\phi}^2+ i
m_{h,\phi}\Gam_{h,\phi}} \,~.
\eeq
Note that $\M_{++}=\M_{--}$ and  $\M_{+-}=\M_{-+}$ are guaranteed
by CP invariance.
In principle, the photon polarization can separate
the contribution of the scalar mediation
from that of KK gravitons even though
leading contribution comes from the KK graviton mediation.

Brief comments on $\gamma\gamma$ colliders are in order here \cite{Ginzburg}.
From the head-on collisions between the laser
and energetic electron (or positron) beams,
high energy photons are produced.
If we denote the fraction of the photon beam energy
to the initial electron beam energy by
$x={E_\gamma}/{E_e}$,
its maximum value is $x_{\rm max}=z/(1+z)$
where $z=4E_e \omega_0/m_e^2$.
Here $E_\gamma$, $E_e$, $\omega_0$ are the photon, electron and laser beam
energies, respectively.
Usually, $z$ is optimized to be $2(1+\sqrt{2})$
to avoid the $e^+ e^-$ pair production
through the interactions of the laser beam and the backward
scattered photon beam.
In the numerical analysis,
we consider the following range for $x_{1,2}$:
\beq
\sqrt{0.4}\le x_{1(2)} \le x_{\rm max}|_{z=2(1+\sqrt{2})}~.
\eeq

With the given polarizations of the laser and parent electron (positron) beams,
their Compton back-scattering leads to
the differential cross section
\begin{eqnarray}
\label{eq:dsigdz-general}
\frac{d\sigma}{d\cos\theta^*}&=&\frac{1}{32\pi s_{ee}}
\int \int dx_1 dx_2 \frac{f(x_1)f(x_2)}{x_1 x_2} \,
\lambda^{1/2}\left(1,\frac{m_h^2}{\hats},\frac{m_\phi^2}{\hats }\right)
\nonumber \\
&&\times \bigg[ \Big( \frac{1+\xi_2(x_1)\xi_2(x_2)}{2}\Big)
\Big|{\cal M}_{J_z=0} \Big|^2
+\Big( \frac{1-\xi_2(x_1)\xi_2(x_2)}{2}\Big)
\Big|{\cal M}_{J_z=2} \Big|^2 \bigg],
\end{eqnarray}
where $s_{ee}=\hats/(x_1x_2)$ is the square of the
c.m. energy of the parent $e^+ e^-$ collision,
and
\begin{eqnarray}
\label{sum}
\Big|{\cal M}_{J_z=0} \Big|^2 &=& \frac{1}{2}\bigg[
\Big|{\cal M}_{++}\Big|^2 + \Big|{\cal M}_{--}\Big|^2 \bigg],
\\ \nonumber
\Big|{\cal M}_{J_z=2} \Big|^2 &=& \frac{1}{2}\bigg[
\Big|{\cal M}_{+-}\Big|^2 + \Big|{\cal M}_{-+}\Big|^2 \bigg].
\end{eqnarray}
Here $f(x)$ is the photon luminosity function and $\xi_2(x)$ is the
averaged circular polarization of the
back-scattered photon beam, both of which depend on
the polarizations of the electron $P_e$
and laser beam $P_l$.
The explicit expressions for $f(x)$ and $\xi_2(x)$ are
\begin{equation}
f(x, P_e, P_l; z) =
\frac{1}{\hat{\sigma}_{_{C}}} C(x)~,
\label{f(x)}
\end{equation}
where
\bea
\hat{\sigma}_{_{C}} &=&
 \left[\left(1 - \frac{4}{z} -\frac{8}{z^2}\right) \ln (z + 1) +
\frac{1}{2} + \frac{8}{z} -
\frac{1}{2 (z + 1)^2}\right]
\\ \nonumber &&
+ P_e \, P_l
\left[\left(1 + \frac{2}{z}\right) \ln (z + 1)
- \frac{5}{2} + \frac{1}{z + 1} - \frac{1}{2 (z + 1)^2}\right].
\label{sigC}
\eea
and
\begin{equation}
C(x) \equiv \frac{1}{1 - x} + (1 - x) - 4 r (1 - r)
- P_e \, P_l \, r \, z (2 r - 1) (2 - x),
\label{C(x)}
\end{equation}
where $r \equiv x/[z(1 - x)]$.
The average helicity $\xi_2(x, P_e, P_l; z)$ is given by
\begin{equation}
\xi_2(x, P_e, P_l; z) = \frac{1}{C(x)}
\left\{P_e \, \left[\frac{x}{1 - x} + x (2 r - 1)^2\right]
- P_l \, (2r - 1)\left(1 - x + \frac{1}{1 - x}\right)\right\}.
\label{xi2}
\end{equation}

\begin{figure}[t!]
\centering
\includegraphics[width=4in]{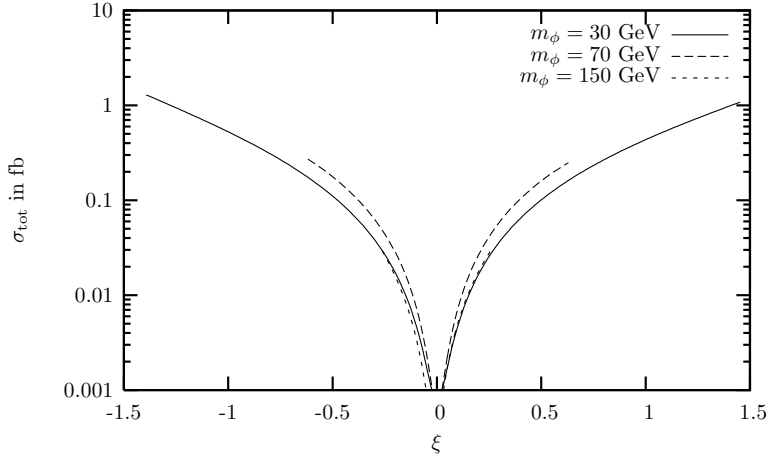}
\caption{\label{fig:totsig_xi} \small
Total cross section for $\gamma\gamma \to h \phi$ in fb
as a function of $\xi$ for $m_\phi=30,~70,~150$ GeV at $e^+ e^-$ colliders
running in the $\gamma\gamma$ mode using laser backscattering.
All of the polarizations are set to be zero,
$m_h =120$ GeV, $\lphi=3.5$ TeV, and $\sqrt{s_{ee}}=500$ GeV.}
\end{figure}

In Fig.~\ref{fig:totsig_xi}, we plot the total cross section as a
function of $\xi$
for $m_\phi=30$, $70$, $150$ GeV.
We set $m_h=120$ GeV, $\lphi=3.5$ TeV and $\sqrt{s_{ee}}=500$ GeV.
Note that the requirement of positive definiteness of the mass and kinetic
terms
limits the range of the mixing parameter $\xi$.
Here we assume that all beams are unpolarized.
For $m_\phi=30$ GeV, the maximum total cross section can reach about 1 fb,
which will produce about 1000 events with 1 ab$^{-1}$ luminosity.
For heavier radion mass, the more restricted $\xi$-range reduces
the maximum of the total cross section, {\it e.g.}, to several
$10^{-2}$ fb for $m_\phi= 150$ GeV.

\begin{figure}[t!]
\centering
\includegraphics[width=4in]{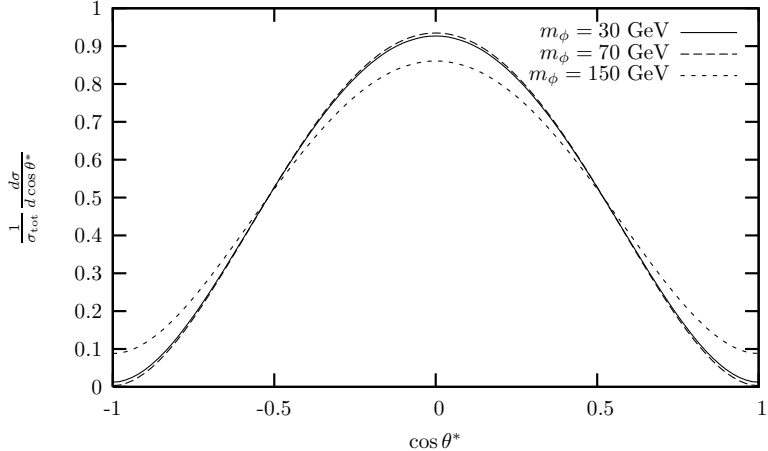}
\caption{\label{fig:dsigdz_totsig} \small
Normalized differential cross section $\frac{1}{\sigma_{\rm tot}}\,
\frac{d \sigma }{d \cos\theta^*}$ for $m_\phi=30,~70,~150$ GeV.
The $m_\phi$ dependence is mild.
All of the polarizations are set to be zero,
$m_h =120$ GeV, $\lphi=3.5$ TeV and $\xi =0.26$.}
\end{figure}

Since the $\gm\gm \to h \phi$ process is practically mediated by the
massive spin-2 KK graviton, the angular distribution shows its
characteristic behavior proportional to $\sin^4\theta^*$, as shown
in Eq.~(\ref{eq:helicityamp+-}). In Fig.~\ref{fig:dsigdz_totsig}, we
plot the normalized differential cross section $(1/\sigma){d \sigma
}/{d \cos\theta^*}$ versus $\cos\theta^*$ for $m_\phi=30,~70,~150$
GeV. Here we have set $\lphi=3.5$ TeV, and $\xi =0.26$ which is the
allowed maximum value for $m_\phi=150$ GeV. For $m_\phi=30$ GeV and
$m_\phi=70$ GeV, the distributions are practically the same,
proportional to $\sin^4\theta^*$. For $m_\phi=150$ GeV, the enhanced
contributions from the Higgs boson and radion alter the behavior
slightly. 
Especially, at the end points ($\cos\theta^* = \pm 1$) the cross section
comes solely from the Higgs and radion exchanges.
One crucial reason is that the heavy radion mass reduces the
KK-graviton contribution which has an additional factor of
$\lambda^2(1,m_h^2/\hats,m_\phi^2/\hats)$ with respect to the
$h/\phi$ contribution, as can be seen from
Eqs.~(\ref{eq:helicityamp++}) and (\ref{eq:helicityamp+-}).

\begin{figure}[t!]
\centering
\includegraphics[width=4in]{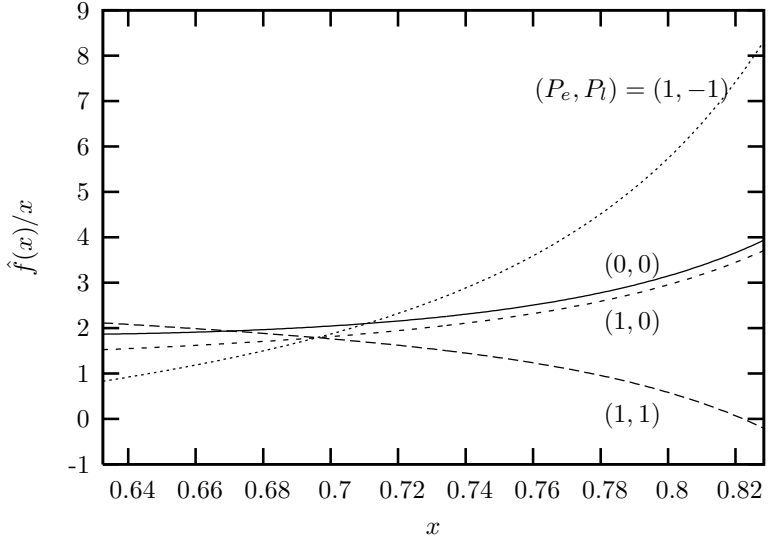}
\caption{\label{fig:photon_pdf} \small
$\hat{f}(x)/x$ versus $x$ for various polarizations of the electron and laser
beams, denoted by $(P_e,P_l)$.
The $\hat{f}(x)=f(x)$ for unpolarized beams $(P_e=P_l=0)$,
and $\hat{f}(x)=f(x)\xi_2(x)$ for polarized beams with $P_e=1$.
The beam polarizations of $(P_e,P_l)=(1,-1)$
generate the highest energy photon beam from the parent electron beam.}
\end{figure}

As the general expression of Eq.~(\ref{eq:dsigdz-general}) suggests,
the appropriate adjustment of the electron (or positron) and laser
beam polarizations can enhance the production rate. As can be seen
from Eq.(\ref{eq:dsigdz-general}), non-zero and negative
$\xi_2(x_1)\xi_2(x_2)$ can enhance the graviton contribution.
Another merit is that the polarized beams can enhance the energy of
back-scattered photon beam and thus increase the signal cross
section. Figure \ref{fig:photon_pdf} presents $\hat{f}(x)/x$ where
$\hat{f}(x)=f(x) \xi_2(x)$ for the polarized beam while
$\hat{f}(x)=f(x)$ for the unpolarized beam\footnote{The average
helicity function $\xi_2(x)$ becomes zero for $P_e=P_l=0$.}.
Compared to the unpolarized beam case, $(P_e,P_l)=(1,-1)$
combination generates a more energetic photon beam. Moreover, the
spin-2 nature of KK graviton prefers opposite polarizations for
$e^+$ and $e^-$ beams. Therefore, the optimal polarizations would be
$(P_{e^-},P_{e^+}, P_{l1}, P_{l2})=(1,-1,-1,1)$.

\begin{figure}[t!]
\centering
\includegraphics[width=4in]{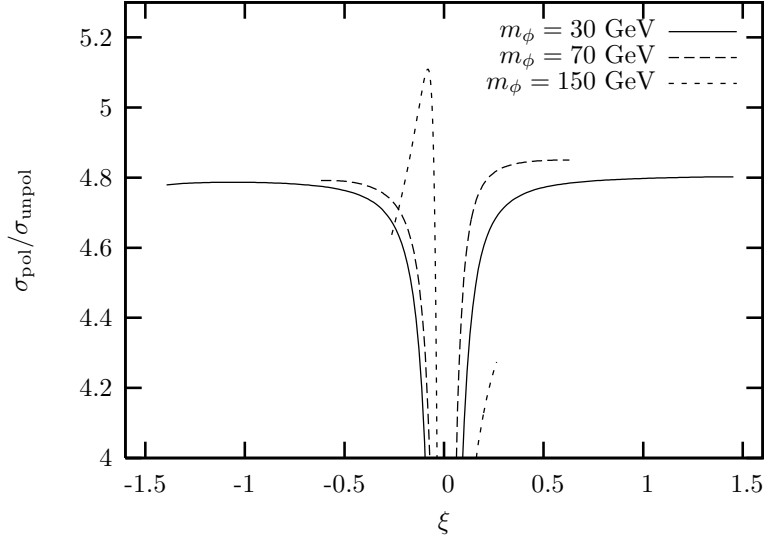}
\caption{\label{fig:pol} \small
The ratio of the polarized total cross section to unpolarized one,
as a function of $\xi$ for $m_\phi=30,~70,~150$ GeV.
The polarization is set to be $(P_{e^-},P_{e^+}, P_{l1}, P_{l2})=(1,-1,-1,1)$,
$m_h=120$ GeV,
$\sqrt{s_{ee}}=500$ GeV, and $\lphi=3.5$ TeV.}
\end{figure}

With the optimal polarization combination, we plot the ratio of the
polarized total cross section to unpolarized one, as a function of
$\xi$ for $m_\phi=30,~70,~150$ GeV in Fig.~\ref{fig:pol}. The
$\xi$-dependence is negligible in the region of $|\xi| \gsim 0.3$. The
enhancement is significant and better than the naive estimation of a
factor of two from Eq.~(\ref{eq:dsigdz-general}). The enhancement
factor can reach upto $5.1$ for $m_\phi=150$ GeV, and about $4.8$
for $m_\phi=30,\,70$ GeV. Even for the $m_\phi=70$ GeV case, the
total cross section can be about 1 fb. Under conservative assumption
for the electron and positron beam polarization, we set
$(P_{e^-},P_{e^+}, P_{l1}, P_{l2})=(0.8,-0.6,-1,1)$: the maximum of
the enhancement factor is about $3.3$ for $m_\phi=30,\,70$ GeV, and
about $3.6$ for $m_\phi=150$ GeV.

\section{Decays and Detection of the radion-Higgs pair}

In this section, we consider the feasibility of detecting $h\phi$
pair production.
For a Higgs boson of mass around $120$ GeV,
the major decay mode is into $b\bar b$.
The partial decay rate into $WW$
will begin to grow at $m_h \agt 140$ GeV.
Therefore, we shall focus on the
$b\bar b$ mode for the Higgs boson decay.
A light radion, on the other hand,
has the major decay mode of $gg$ because of the QCD trace anomaly,
followed by $b\bar b$ (a distant second).
When the radion mass
gets above the $WW$ threshold, the $WW$ mode becomes dominant.

The major background comes from the QCD heavy-flavor production of
\begin{equation}
\gamma\gamma \to b\bar b/c\bar c + 2\; {\rm jets}~,
\end{equation}
where each jet can come from a gluon or a light quark.
Here the $c\bar c$ pair can also fake the $B$-tag though with a much lower
probability than the $b$ quark.
We calculate the QCD $b\bar b/c\bar c + 2$ jets background by a
parton-level calculation, in which the sub-processes are generated by
MADGRAPH\,\cite{madgraph}.
Another possible source of background is
\begin{equation}
\gamma\gamma\to W^+ W^-\,,
\end{equation}
followed by the hadronic decay of the $WW$ pair.
The decay of the $W$ boson into a $b$ quark is severely suppressed by
$(V_{cb}+V_{ub})^2 \simeq (0.05)^2 $.
The chance of seeing two $b$ quarks in the $WW$ decay is very rare,
of the order of $6\times 10^{-6}$.
On the other hand, $WW$ production is still a possible background because of
the $W$ boson decay into a $c$ quark.  The charm
quark may be tagged with a displaced vertex with a small mistag
probability, thus may be misinterpreted as a $b$ quark.

We assume a 50\% $B$-tagging efficiency and a chance of 5\% mistag
(a charm quark misinterpreted as a $b$ quark) in our study.
Typical cuts on detecting
the $b$-jets and light jets are applied:
\begin{eqnarray}
p_T(b) > 15 \;{\rm GeV}, &&\; p_T(j) > 15 \; {\rm GeV}, \nonumber \\
|\cos \theta_b |< 0.9,\; |\cos \theta_j|<0.9\;, &&
\cos \theta_{b,\bar b},\,
\cos \theta_{j j},\,
\cos \theta_{b, j}  < 0.9, \;\;  \nonumber
\end{eqnarray}
where $\cos\theta_b$ and $\cos\theta_j$ denote the cosine of the angle
of the outgoing $b$ quark and the jet, respectively,
and
$\cos \theta_{b, \bar b}$, $\cos\theta_{j j}$,
and $\cos\theta_{b,j}$ denote the cosine of the angle between
the two $b$ quarks, between the two jets, and between the $b$ quark and
the jet, respectively.
The angular cuts are mainly for the detection purpose and for removing
the collinear divergence in the calculation.
We have applied a gaussian smearing
$\Delta E/E = 0.5/\sqrt{E}$, where $E$ is in GeV, to the final-state
$b$-jets and light jets to simulate the detector resolution.
Since the Higgs boson is produced together with
a radion mainly via an intermediate graviton KK state,
the Higgs boson tends
to have a large $p_T \sim m_{G^{(1)}}/2$.
Therefore, a
transverse momentum cut on the $b\bar b$ pair is
very efficient against the QCD background while only hurts the signal
marginally.
We apply a cut
\begin{equation}
\label{eq:ptcut}
p_T(b\bar b) > 100 \;{\rm GeV}~,
\end{equation}
to reduce the background.
Finally, we apply the invariant mass constraint of the $b\bar b$ pair to
be near the Higgs boson mass and that of the jet pair to be near
the radion mass:
\begin{equation}
\left| m_{b\bar b} - m_H \right | < 10 \;{\rm GeV}\;, \qquad
\left| m_{jj} - m_\phi \right | < 10 \;{\rm GeV}\;.
\end{equation}

\begin{table}[t!]
\caption{ \small \label{table1}
Cross sections in fb under successive application of the cuts mentioned in the
text at $\sqrt{s_{ee}}=0.5$ TeV.
Unpolarized beams are chosen.
The results for $m_\phi =30\; (\xi=1.4)$, $70\; (\xi=0.6)$
and $150\; (\xi=0.25)$ GeV are shown in the corresponding rows.
}
\medskip
\begin{ruledtabular}
\begin{tabular}{ccccc}
Cuts &  $h\phi$ signal & $b\bar b jj$ & $c \bar c jj$ & $W^+ W^- \to c\bar c
jj$ \\
\hline
$p_T(b,j)>15$  GeV              &  0.14   &  & & \\
$|\cos\theta_{b,\bar b,j}|<0.9$ & (0.16) & 39  & 346  &  287  \\
$\cos\theta_{b_i j} < 0.9$      & (0.018) & & & \\
\hline
additional             &  0.14  & & & \\
$p_T(b\bar b)>100$ GeV & (0.15) &  4.5   & 38  & 26 \\
                       & (0.016) &        &      &  \\
\hline
additional                     &  0.14  & 0.092 & 1.32  & 0.0009 \\
$|M_{b\bar b} - m_H| < 10$ GeV &  (0.15)  & (0.091) & (1.0)& (0.242)\\
$|M_{j\bar j} - m_\phi| < 10$ GeV&(0.014) & (0.027)  & (0.215) &  (0.82) \\
\hline
with B-tag $\epsilon_b=0.5$   & 0.034& 0.023   & 0.0033 &$2\times 10^{-6}$\\
or C-mistag $\epsilon_c=0.05$ &(0.037)& (0.023) &(0.0025)&($6\times 10^{-4}$)\\
         & (0.0035)  & (0.007) & ($5\times 10^{-4}$) &  ($2\times 10^{-3}$) \\
\end{tabular}
\end{ruledtabular}
\end{table}

We summarize the cross sections for
the signal and various backgrounds under successive cuts
in Table \ref{table1}.
The final signal-to-background ratio is quite promising.  For $m_\phi=30$ GeV
and $\xi=1.4$ we obtain a signal-to-background ratio about $1.3:1$.  For
$m_\phi=70$ GeV with $\xi=0.6$ a ratio of $1.4:1$
can be obtained.  A ratio of $0.37:1$ is obtained for the case of
$m_\phi=150$ GeV and $\xi=0.25$.  Note that the signal cross section scales
as $\xi^2$ as long as the positive mass-square
constraints for the scalar bosons are satisfied.

\begin{table}[t!]
\caption{ \small \label{table2}
Cross sections in fb under successive application of the cuts mentioned in the
text at $\sqrt{s_{ee}}=0.8$ TeV.
Unpolarized beams are chosen. The mixing parameter $\xi=1.4$
and $m_\phi=30$ GeV
are chosen. }
\medskip
\begin{ruledtabular}
\begin{tabular}{ccccc}
Cuts &  $h\phi$ signal & $b\bar b jj$ & $c \bar c jj$ & $W^+ W^- \to c\bar c
jj$ \\
\hline
$p_T(b,j)>15$  GeV &   &  & & \\
$|\cos\theta_{b,\bar b,j}|<0.9$ & 6.25  & 31  & 280   &  156  \\
$\cos\theta_{b_i j} < 0.9$      &  & & & \\
\hline
additional & & & & \\
$p_T(b\bar b)>200$ GeV & 6.10 & 0.86  & 7.9 & 3.0 \\
\hline
additional & & & & \\
$|M_{b\bar b} - m_H| < 10$ GeV & & & & \\
$|M_{j\bar j} - m_\phi| < 10$ GeV & 5.95 & 0.0015 & 0.031 & $\sim 0$  \\
\hline
with B-tag $\epsilon_b=0.5$  & & & & \\
or C-mistag $\epsilon_c=0.05$ & 1.49 & 0.0004 & $8\times 10^{-5}$ & $\sim 0$
\end{tabular}
\end{ruledtabular}
\end{table}

In Table \ref{table2} we show the results at $\sqrt{s_{ee}}=0.8$ TeV, at
which the effect of the first graviton resonance is large, such that the
signal cross section is substantially larger than the background after the
cuts.  This is because the majority of the photon collisions are
at $\sqrt{\hat s} = \sqrt{ x_1 x_2 s_{ee}} \sim (0.7-0.8)(800\,{\rm GeV})$,
which is very close to the first KK graviton resonance.


To disentangle possible conflict between the electroweak precision
data and the direct search bound of the SM Higgs, we studied the
possibility that
the SM Higgs mixes with the radion of the RS model such that the Higgs
branching ratio
into $b \bar b$ becomes smaller and thus escapes the limit of direct search.
We have explored the direct search of the radion-Higgs associated production
at photon colliders, and
shown that the photon colliders achieved by the laser backscattering
technique are very special in probing such a process.
As is well known, the advantage of photon colliders is the capability
of adjusting
the polarization of incoming photons such that the signal of the
spin-2 graviton
exchange can be largely enhanced compared to the unpolarized collision.
We have found the enhancement factor is around five, unless $\xi$ is quite small.
This can be attained by aligning the polarization of the incoming photons
such that the nature of the spin-2 graviton exchange is fully enhanced.
We have also studied the corresponding background to $h \phi$ pair
productions at photon colliders step-by-step in detail.
Specifically,
we have considered the major backgrounds from the QCD heavy flavor
production of $\gamma \gamma \to b \bar b/c \bar c + 2$ jets and
$\gamma \gamma \to W^+ W^- \to c \bar c + 2$ jets.
By imposing the various cuts, we have shown that the associated
production signal of
$h \phi$ can be comparable or even much larger than those
backgrounds at photon colliders.

\begin{acknowledgments}

\noindent K.C. was supported by the NSC of Taiwan under
Grant Nos.\ NSC 93-2112-M-007-025- and 94-2112-M-007-010-.
The work of C.S.K. was supported
in part by  CHEP-SRC Program and
in part by the Korea Research Foundation Grant funded by the Korean Government
(MOEHRD) No. R02-2003-000-10050-0.
The work of JS is supported by KRF under grant No. R04-2004-000-10164-0.

\end{acknowledgments}

\end{document}